\documentclass[10pt,onecolumn]{IEEEtran}

\usepackage{amsmath,amssymb,amsfonts,graphicx}
\usepackage{latexsym}
\usepackage{amsthm}
\usepackage{cite}

\usepackage{algorithm}
\usepackage{algpseudocode}

\usepackage{caption}
\usepackage{subcaption}

\newcommand{\database}{{\mathcal{D}^n}}    % randomized mechanism
\newcommand{\noise}{{\mathbf{X}}}    %  symbol denoting the noise
\newcommand{\KM}{{\mathcal{K}}}    % randomized mechanism
\newcommand{\e}{{\epsilon}}    % epsilon
\newcommand{\loss}{{\mathcal{L}}}    % cost function / loss function
\newcommand{\D}{{\Delta}}    % query sensitivity
\newcommand{\p}{{\mathcal{P}}}  % probability measure

\newcommand{\sP}{{\mathcal{SP}}}  % set of probability measures satisfying differential privacy constraint
\newcommand{\psym}{{\p_{\text{sym} }}}
\newcommand{\sPsymi}{{\sP_{i, \text{sym} }}}

\newcommand{\dd}{{\D'}}

\newcommand{\R}{{\mathbb{R}}}  % real number
\newcommand{\N}{{\mathbb{N}}}  % natural number, including 0
\newcommand{\Z}{{\mathbb{Z}}}  % integer number

\newcommand{\pa}{{ \sP_{i, \text{md}}  }}  % first sequence \p_i,  monotone decreasing
\newcommand{\pb}{{ \sP_{i, \text{pd}}  }}  % second sequence \p_i, periodically decayig
\newcommand{\pc}{{ \sP_{i, \text{fr}}  }}  % third sequence \p_i, first and last ratio is e^\e (fr = fixed ratio)
\newcommand{\pd}{{ \sP_{i, \text{step}} }}  % fourth sequence \p_i, step function, except something in between

\newcommand{\nx}{{x}}  %  may only use it once

\newcommand{\bx}{{\mathbf{x}}}  % mathbold x
\newcommand{\by}{{\mathbf{y}}}  % mathbold x

\newcommand{\bt}{{\mathbf{t}}}  % mathbold t

\newcommand{\cost}{{\mathcal{C}}}  %  single variable cost function
\newtheorem{theorem}{Theorem}
\newtheorem{lemma}[theorem]{Lemma}
\newtheorem{definition}{Definition}
\newtheorem{corollary}[theorem]{Corollary}

\begin{document}

\title{The Optimal Mechanism in Differential Privacy: Multidimensional Setting}

\author{
\authorblockN{Quan Geng, and Pramod Viswanath}\\
\authorblockA{Coordinated Science Laboratory and Dept. of ECE \\
University of Illinois, Urbana-Champaign, IL 61801 \\
Email: \{geng5, pramodv\}@illinois.edu} }

\maketitle

\begin{abstract}
We derive the optimal $\epsilon$-differentially private mechanism for a general two-dimensional real-valued (histogram-like) query function under a utility-maximization (or cost-minimization) framework for the $\ell^1$ cost function. We show that the optimal noise probability distribution has a correlated multidimensional staircase-shaped probability density function. Compared with the Laplacian mechanism, we show that in the high privacy regime (as $\epsilon \to 0$), the Laplacian mechanism is approximately optimal; and in the low privacy regime (as $\epsilon \to +\infty$), the optimal cost is $\Theta(e^{-\frac{\epsilon}{3}})$, while the cost of the Laplacian mechanism is $\frac{2\Delta}{\epsilon}$, where $\Delta$ is the sensitivity of the query function. We conclude that the gain is more pronounced in the low privacy regime. We conjecture that the optimality of the staircase mechanism holds for vector-valued (histogram-like) query functions with arbitrary dimension, and holds for many other classes of cost functions as well.

\end{abstract}

\section{Introduction} \label{sec:intro}
Differential privacy is a framework to quantify to what extent  individual privacy in a statistical database is preserved while releasing useful statistical information about the database \cite{DMNS06}. The basic idea of differential privacy is that the presence of any individual data in the database should not affect the final released statistical information   significantly, and thus it can give strong privacy guarantees against an adversary with arbitrary auxiliary information. For more background  and motivation of differential privacy, we refer the readers to the survey by Dwork \cite{DPsurvey}.

The standard approach to preserving differential privacy for real-valued query function is to perturb the query output by adding random noise with the Laplacian distribution. Recently, Geng and Viswanath \cite{GV13} show that under a general utility-maximization framework, for single real-valued query function, the optimal $\e$-differentially private mechanism is the staircase mechanism, which adds noise with the staircase distribution to the query output.

In this work, we study the optimal mechanism in $\e$-differential privacy for a vector-valued (histogram-like) query function in the multiple dimensional setting, where the query output has multiple components, each of which is real-valued, and the global sensitivity of the query function is defined using the $\ell^1$ metric. For instance, the histogram function is a vector-valued query function with global sensitivity equal to one, and it has been widely studied in the literature, e.g., \cite{Fang12,Hay10,geometry,Xu2012,Li10,NTZ12}.
 We extend the optimality of the staircase mechanism from the single dimensional setting to the multiple dimensional setting.  We show that when the dimension of the query output is two, for the $\ell^1$ cost function, the optimal query-output independent perturbation mechanism is to add noise with a staircase-shaped probability density function to the query output.  Compared with the Laplacian mechanism, we show that in the high privacy regime (as $\epsilon \to 0$), the Laplacian mechanism is approximately optimal; and in the low privacy regime (as $\epsilon \to +\infty$),  the optimal cost is $\Theta(e^{-\frac{\epsilon}{3}})$, while the cost of the Laplacian mechanism is $\frac{2\D}{\epsilon}$. We conclude that the gain is more pronounced in the low privacy regime. We conjecture that the optimality of the staircase mechanism holds for vector-valued (histogram-like) query functions with arbitrary dimension, and holds for many other classes of cost functions as well. We discuss how to make progress on proving this conjecture at the end of Section \ref{sec:result}.

It is natural to compare the performance of the optimal multiple dimensional (correlated) staircase mechanism and the composite single-dimensional staircase mechanism \cite{GV13},  which adds independent staircase noise to each component of the query output. In the context of a two-dimensional query function, if independent staircase noise is added to each component of query output, to satisfy the $\e$-differential privacy constraint, the parameter of the  staircase noise  is $\frac{\e}{2}$ instead of $\e$, and thus the total cost will be proportional to $e^{-\frac{\e}{4}}$, which is worse than the optimal cost $\Theta(e^{-\frac{\e}{3}})$.

\subsection{Related Work}

Dwork et al. \cite{DMNS06} introduce $\e$-differential privacy and show that the Laplacian mechanism, which perturbs the query output by adding random noise with Laplace distribution proportional to the global sensitivity of the query function, can preserve $\e$-differential privacy. Indeed, the query functions studied in \cite{DMNS06} are histogram-like functions, and the global sensitivity is defined using the $\ell^1$ metric.
Nissim, Raskhodnikova and Smith \cite{NRS07} show that for certain nonlinear query functions, one can improve the accuracy by adding data-dependent noise calibrated to the smooth sensitivity of the query function, which is based on the local sensitivity of the query function.
McSherry and Talwar \cite{McSherry07} introduce the \emph{exponential mechanism} to preserve $\e$-differential privacy for general query functions in an abstract setting, where the query function may not be real-valued.
Dwork et al. \cite{DKMMN06} introduce $(\e,\delta)$-differential privacy and show that adding random noise with Gaussian distribution can preserve $(\e,\delta)$-differential privacy for real-valued query function.

Kasiviswanathan and Smith \cite{Smith08} study $(\e,\delta)$-semantic privacy under a Bayesian framework. Chaudhuri and Mishra \cite{Chaudhuri06}, and Machanavajjhala et al. \cite{Kifer08} propose different variants of the standard $\e,\delta$-differential privacy.

Ghosh, Roughgarden, and Sundararajan  \cite{Ghosh09} show that for a single count query with sensitivity $\D = 1$, for a general class of utility functions, to minimize the expected cost under a Bayesian framework the optimal mechanism to preserve $\e$-differential privacy is the geometric mechanism, which adds noise with geometric distribution. Brenner and Nissim  \cite{Nissim10}  show that for general query functions no universally optimal mechanisms exist. Gupte and Sundararajan \cite{minimax10} derive the optimal noise probability distributions for a single count query with sensitivity $\D = 1$ for minimax (risk-averse) users. \cite{minimax10} shows that although there is no universally optimal solution to the minimax optimization problem in \cite{minimax10} for a general class of cost functions, each solution (corresponding to different cost functions) can be derived from the same geometric mechanism by randomly remapping. Geng and Viswanath \cite{GV13} generalize the results of \cite{Ghosh09} and \cite{minimax10} to real-valued (and integer-valued) query functions with arbitrary sensitivity, and show that the optimal query-output independent perturbation mechanism is the staircase mechanism, which adds noise with a staircase-shaped probability density function (or probability mass function for integer-valued query function) to the query output. Geng and Viswanath \cite{GV13} show that under a general utility-maximization framework, for single real-valued query function, the optimal $\e$-differentially private mechanism is the staircase mechanism, which adds noise with the staircase distribution to the query output.

Differential privacy for histogram query functions has been widely studied in the literature, e.g.,  \cite{Fang12,Hay10,geometry,Xu2012,Li10,NTZ12}, and many existing works use the Laplacian mechanism as the basic tool. For instance, Li et al. \cite{Li10} introduce the matrix mechanism to answer batches of linear queries over a histogram in a differentially private way with good accuracy guarantees. Their approach is that instead of adding Laplacian noise to the workload query output directly, the matrix mechanism will design an observation matrix which is the input to the database,  from perturbed output (using the standard Laplace mechanism)  estimate the histogram itself, and then compute the query output directly.  \cite{Li10} shows that this two-stage process will preserve differential privacy and increase the accuracy. Hay et al. \cite{Hay10} show that for a general class of histogram queries, by exploiting the consistency constraints on the query output, which is differentially private by adding independent Laplace noises, one can improve the accuracy while still satisfying differential privacy. These existing works study how to efficiently answer a set of linear queries on the histogram, while our work addresses the problem of releasing the histogram itself, which can be viewed as the worst-case query release (without knowing which linear queres will be asked).

Hardt and Talwar \cite{geometry} study the tradeoff between privacy and error for answering a set of linear queries over a histogram under $\epsilon$-differental privacy. The  error is defined as the worst  expectation of the $\ell^2$-norm of the noise. \cite{geometry} derives a lower bound for the error in the high privacy regime by using tools from convex geometry and Markov's inequality, and gives an upper bound by analyzing a differentially private mechanism, $K$-norm mechanism, which is an instantiation of the exponential mechanism and involves randomly sampling from a high dimensional convex body.
Nikolov, Talwar and Zhang \cite{NTZ12} extend the result of \cite{geometry} on answering linear querys over a histogram to the case of  $(\e,\delta)$-differential privacy. Using tools from discrepancy theory, convex geometry and statistical estimation, they derive lower bounds and upper bounds of the error, which are within a multiplicative factor of $O(\log \frac{1}{\delta})$ in terms of $\delta$.

Kasiviswanathan, Rudelson, Smith and Ullman \cite{KRSU10} derive lower bounds on the noise for releasing contingency tables under $(\e,\delta)$-differential privacy constraint. Anindya De \cite{De12}  studies lower  bound on the additive noise for Lipschitz query functions in $(\e,\delta)$-differential privacy which uses a different metric for the noise, and the lower bound depends on the size of the database.

\subsection{Organization}

This paper is organized as follows. We formulate the utility-maximization/cost-minimization under the $\e$-differential privacy constraint as a functional optimization problem in Section \ref{sec:formulation}. We present our main result on the optimality of multiple dimensional (correlated) staircase mechanism in Section \ref{sec:result}. Section \ref{sec:gamma} studies the asymptotic properties and performances of the correlated staircase mechanism for the $\ell^1$ cost function. Section \ref{sec:proof} gives a detailed proof of our main result Theorem \ref{thm:main}.

\section{Problem Formulation} \label{sec:formulation}

Consider a multidimensional real-valued query function
\begin{align}
	q: \database \rightarrow \R^d,
\end{align}
where $\database$ is the domain of the databases, and $d$ is the dimension of the query output. Given $D \in \database$, the query output can be written as
\begin{align}
 	q(D) = (q_1(D), q_2(D), \dots, q_d(D)),
 \end{align}
 where $q_i(D) \in \R, \forall 1 \le i \le d $.

The global sensitivity of the query function $q$ is defined as
\begin{align}
	\D \triangleq \max_{ D_1,D_2 \subseteq \database: |D_1 - D_2| \le 1} \| q(D_1) - q(D_2)\|_1 = \sum_{i=1}^d |q_i(D_1) - q_i(D_2)|, \label{def:sensitivity}
\end{align}
where the maximum is taken over all possible pairs of neighboring database entries $D_1$ and $D_2$ which differ in at most one element, i.e., one is a proper subset of the other and the larger database contains just one additional element \cite{DPsurvey}. For instance, the global sensitivity of a histogram query function is one, since each element in the dataset can affect only one component of the query output by one.

\begin{definition}[$\e$-differential privacy \cite{DPsurvey}]
	A randomized mechanism $\KM$ gives $\e$-differential privacy if for all data sets $D_1$ and $D_2$ differing on at most one element, and all $S \subset \text{Range}(\KM)$,
	\begin{align}
	 	\text{Pr}[\KM(D_1) \in S] \le e^{\e} \;  \text{Pr}[\KM(D_2) \in S]. \label{eqn:dpgeneral}
	 \end{align}
\end{definition}

The standard approach to preserving the differential privacy is to add noise to the output of query function. Let $q(D)$ be the value of the query function evaluated at $D \subseteq \database$, the noise-adding  mechanism $\KM$ will output
\begin{align}
 	\KM(D) = q(D) + \noise = (q_1(D) + X_1, \dots, q_d(D) + X_d),
 \end{align}
where $\noise = (X_1, \dots, X_d) \in \R^d$ is the noise added by the mechanism to the output of query function. Due to the optimality of  query-output independent perturbation mechanisms (under a technical condition) in \cite{GV13}, in this work we restrict ourselves to query-output independent noise-adding mechanisms, i.e., we assume that the noise $\noise$ is independent of the query output.

In the following we derive the differential privacy constraint on the probability distribution of $\noise$ from \eqref{eqn:dpgeneral}.
\begin{alignat}{2}
	 & \; \text{Pr}[\KM(D_1) \in S] & &\le  e^{\e}\; \text{Pr}[\KM(D_2) \in S] \\
	 \Leftrightarrow & \; \text{Pr}[ q(D_1) + \noise \in S]  & &\le e^{\e} \; \text{Pr}[q(D_2) + \noise \in S] \\
	 \Leftrightarrow & \; \text{Pr}[  \noise \in S - q(D_1) ]  & &\le e^{\e} \; \text{Pr}[  \noise \in S - q(D_2) ] \\
	 \Leftrightarrow & \; \text{Pr}[  \noise \in S' ] & &\le e^{\e} \; \text{Pr}[  \noise \in S' + q(D_1)- q(D_2) ],  \label{eqn:dpconstraint2}
\end{alignat}
where $S' \triangleq S - q(D_1) = \{ s-q(D_1) | s \in S \}$.

%%%%%%%%%%%%%%%%%%%%% define S + d %%%%%%%%%%%%%%%%%%%%%

Since \eqref{eqn:dpgeneral} holds for all measurable sets $S \subseteq \R^d$, and $\|q(D_1)- q(D_2)\|_1 \le \D$, from \eqref{eqn:dpconstraint2} we have
\begin{align}
	\text{Pr}[  \noise \in S' ] \le e^{\e} \; \text{Pr}[  \noise \in S' + \bt ], \label{eqn:dpconstraint3}
\end{align}
for all measurable sets $S' \subseteq \R$ and for all $ \bt \in \R^d $ such that $\|\mathbf{t}\|_1 \le \D$.

Consider a cost function $\loss(\cdot): \R^d \to \R$ which is a function of the added noise $\noise$. Our goal is to minimize the expectation of the cost subject to the $\e$-differential privacy constraint \eqref{eqn:dpconstraint3}.

More precisely, let $\p$ denote the probability distribution of $\noise$ and use $\p(S)$ denote the probability $\text{Pr}[\noise \in S]$. The optimization problem we study in this paper is

\begin{align}
	\mathop{\text{minimize}}\limits_{\p} & \ \int \int \dots \int_{ \R^d} \loss(\nx_1,\nx_2,\dots,\nx_d) \p (d x_1 d x_2 \dots d x_d) \\
	\text{subject to} & \ \p(S) \le e^{\e} \p(S + \bt), \forall \ \text{measurable set} \ S \subseteq \R^d, \ \forall \| \bt \|_1 \le \D.  \label{eqn:dpconstraintfinal}
\end{align}

We solve the above functional optimization problem and derive the optimal noise probability distribution for $\loss(x_1,\dots,x_d) = \sum_{i=1}^d |x_i|$ with $d=2$.

\section{Main Result} \label{sec:result}
In this section we state our main result Theorem \ref{thm:main}. The detailed proof is given in Section \ref{sec:proof}.

In this work we consider the $\ell^1$ cost function:
\begin{align}
	\loss(x_1,x_2,\dots,x_d) = \sum_{i=1}^d |x_i|, \forall (x_1,x_2,\dots,x_d) \in \R^d.
\end{align}

% \begin{figure}[t]
% \centering
% \includegraphics[scale=0.4]{fgamma.pdf}
% \caption{The Staircase-Shaped Probability Density Function $f_{\gamma}(x)$}
% \label{fig:fgamma}
% \end{figure}

Consider a class of multidimensional probability distributions with symmetric and  staircase-shaped probability density function defined as follows. Given $\gamma \in [0,1]$,   define $\p_{\gamma}$ as the probability distribution with probability density function $f_{\gamma}(\cdot)$ defined as
\begin{align}
f_{\gamma}(\bx)  =
\begin{cases}
  	 e^{-k \e} a(\gamma)  &  \|\bx\|_1 \in [ k \D, (k+\gamma) \D) \; \text{for} \; k \in \N  \\
     e^{-(k+1) \e} a(\gamma)  & \|\bx\|_1 \in [ (k+\gamma) \D,   (k+1)\D) \; \text{for} \; k \in \N
  \end{cases}\label{eqn:deffgamma}
\end{align}
where $a(\gamma)$ is the normalization factor to make
\begin{align}
	\int \int \dots \int_{\R^d} f_{\gamma} (\bx) d x_1 d x_2 \dots d x_d = 1.
\end{align}

Define $b \triangleq e^{-\e}$, and define
\begin{align}
	c_k \triangleq \sum_{i=0}^{+\infty} i^k b^i, \forall k \in \N,
\end{align}
where by convention $0^0$ is defined as 1. Then the closed-form expression for $a(\gamma)$ is
\begin{align}
	a(\gamma) \triangleq \frac{d!}  {2^d \D^d   \sum_{k=1}^d \binom{d}{k}c_{d-k} (b + (1-b)\gamma^k)}.
\end{align}

\begin{figure}[t]
\centering
\includegraphics[scale=0.25]{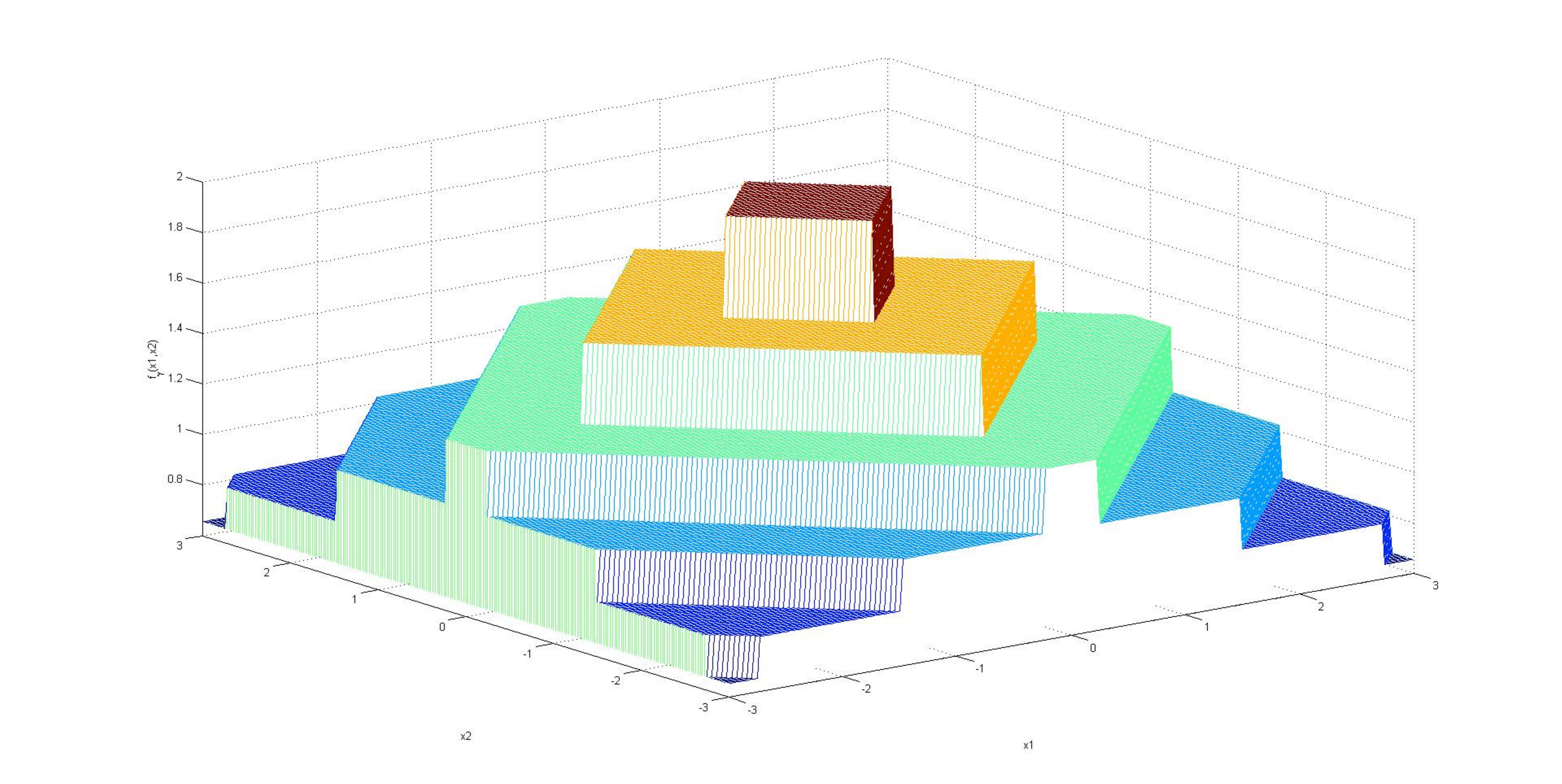}
\caption{Multi-Dimensional Staircase-Shaped Probability Density Function}
\label{fig:staircase2dim}
\end{figure}

It is straightforward to verify that $f_{\gamma}(\cdot)$ is a valid probability density function and $\p_{\gamma}$ satisfies the differential privacy constraint \eqref{eqn:dpconstraintfinal}. Indeed, the probability density function $f_{\gamma}(x)$ satisfies
\begin{align}
	f_{\gamma}(\bx) \le e^{\e} f_{\gamma}(\bx + \bt), \forall \bx \in \R^d, \forall \bt \in \R^d \; \text{s.t.} \; \|\bt \|_1 \le \D,
\end{align}
which implies \eqref{eqn:dpconstraintfinal}.

We plot the probability density function $f_{\gamma}(\bx)$  in Figure \ref{fig:staircase2dim} for $d = 2$. It is easy to see that  $f_{\gamma}(\bx)$ is multi-dimensional staircase-shaped.

Let $\sP$ be the set of all probability distributions which satisfy the differential privacy constraint \eqref{eqn:dpconstraintfinal}.
Our main result is
\begin{theorem}\label{thm:main}
For $d=2$ and the cost function $\loss(\bx) = \|\bx\|_1, \forall \bx \in \R^2$, then
\begin{align}
	\inf_{\p \in \sP}  \int  \int_{\R^2} \loss(\bx) \p (d x_1 d x_2 )  = \inf_{\gamma \in [0,1]}  \int  \int_{ \R^2} \loss(\bx)     f_{\gamma}(\bx)  d x_1 d x_2  .
\end{align}
\end{theorem}

\begin{IEEEproof}
Here we briefly discuss the main proof idea and technique. For the complete proof, see Section \ref{sec:proof}. First, by using a combinatorial argument, we show that given any noise probability distribution satisfying the $\epsilon$-differential privacy constraint, we can discretize the probability distribution by averaging it over each $\ell^1$ layer without increasing the cost. Therefore, we only need to consider those probability distributions with the probability density function being a piecewise constant function of the $\ell^1$-norm of the noise. Second, we show that to minimize the cost, the probability density function as a function of the $\ell^1$-norm of the noise should be monotonically and geometrically
decaying. Lastly, we show that the optimal probability density function should be staircase-shaped.
\end{IEEEproof}

Therefore, the optimal noise probability distribution to preserve $\e$-differential privacy for multidimensional real-valued query function has a staircase-shaped probability density function, which is specified by three parameters $\e$, $\D$ and $\gamma^* = \mathop{\arg \min} \limits_{\gamma \in [0,1] } \int  \int_{ \R^2} \loss(x_1,x_2)     f_{\gamma}(\bx)  d x_1 d x_2$.

We conjecture that Theorem \ref{thm:main} holds for arbitrary dimension $d$. To prove this conjecture, one can reuse the whole proof in Section \ref{sec:proof} and only needs to prove that Lemma \ref{lem:approx} and Lemma \ref{lem:constline} hold for arbitrary $d$, which we believe are true. Lemma \ref{lem:approx} shows that when $d=2$, we can discretize the probability distribution by averaging it over each $\ell^1$ layer without increasing the cost, and the new probability distribution also satisfies the differential privacy constraint. We give a constructive combinatorial argument to prove Lemma \ref{lem:approx} for $d=2$, and believe it holds for arbitrary $d \ge 2$. We prove Lemma \ref{lem:constline} for $d=2$ by studying the monotonicity of the ratio between the cost and volume over each $\ell^1$ layer. Indeed, to prove Lemma \ref{lem:constline}, one only needs to show that $h_k$, which is defined in \eqref{eqn:h}, first decreases and then increases as a function of $k$, and $h_0 \le h_{i-1}$. For fixed $d$, one can derive the explicit formula for $d$ and verify whether $h_k$ satisfies this property (we show it is true for $d=2$ in our proof).

We also conjecture that Theorem \ref{thm:main} holds for other classes of cost functions, which may not be a function only depending on the $\ell^1$-norm of the noise. Numeric simulations suggest that for $d=2$, the correlated multidimensional staircase mechanism is optimal for  $\loss(\bx) = \|\bx\|_2^2$. To prove this conjecture, one has to use a different proof technique, as Lemma  \ref{lem:approx} in our proof does not work for the cost functions which does not depend on the $\ell^1$-norm of the noise only.

%and holds for other classes of cost functions, e.g., $\loss(\bx) = \|\bx\|_2^2$.

\section{Optimal $\gamma^*$ and Asymptotic Analysis} \label{sec:gamma}

In this section, we study the asymptotic properties and performances of the correlated staircase mechanism for the $\ell^1$ cost function.

Note that the closed-form expressions for $c_0, c_1$ and $c_2$ are
\begin{align}
	c_0 &= \frac{1}{1 - b}, \\
	c_1 &= \frac{b}{(1-b)^2},\\
	c_2 &= \frac{b^2+b}{(1-b)^3}.
\end{align}

For $d=2$, we have
\begin{align}
	a(\gamma) &= \frac{1}{2\D^2 \left(  2c_1 (b+(1-b)\gamma) + c_0 (b + (1-b)\gamma^2 )  \right)} \\
			  &= \frac{1}{2\D^2 \left(  \gamma^2 + \frac{2b}{1-b}\gamma + \frac{b+b^2}{(1-b)^2}  \right)}.
\end{align}

Given the two-dimensional staircase-shaped probability density function $f_\gamma(\bx)$, the cost is
\begin{align}
	V(\p_\gamma) &\triangleq \int \int_{\R^2} (|x_1|+|x_2|) f_\gamma(x_1,x_2) \p(d x_1 d x_2 ) \\
	&= 4 \left( \sum_{i=0}^{+\infty}   \int_{i\D}^{(i+\gamma)\D} t t a(\gamma)e^{-i\e} d t +   \sum_{i=0}^{+\infty}   \int_{(i+\gamma)\D}^{(i+1)\D} t t a(\gamma)e^{-(i+1)\e} d t   \right) \\
	&= \frac{4a(\gamma)\D^3}{3} \left(    \sum_{i=0}^{+\infty} b^i (3i^2\gamma + 3i\gamma^2 + \gamma^3) + b   \sum_{i=0}^{+\infty} b^i (3i^2+3i+1 - 3i^2\gamma - 3i \gamma^2 - \gamma^3)   \right) \\
	&= \frac{4a(\gamma)\D^3}{3} \left( 3c_2 \gamma + 3c_1 \gamma^2 + c_0 \gamma^3 + b (3(1-\gamma)c_2 + 3(1-\gamma^2)c_1 + (1-\gamma^3)c_0)   \right) \\
	&= \frac{2\D}{3} \frac{  3c_2 \gamma + 3c_1 \gamma^2 + c_0 \gamma^3 + b (3(1-\gamma)c_2 + 3(1-\gamma^2)c_1 + (1-\gamma^3)c_0)   }{ \gamma^2 + \frac{2b}{1-b}\gamma + \frac{b+b^2}{(1-b)^2}   } \\
	&= \frac{2\D}{3} \frac{ c_0(1-b)\gamma^3 +  3c_1 (1-b)\gamma^2 + 3c_2 (1-b)\gamma   + b (c_0 + 3c_1 + 3c_2)  }{  \gamma^2 + \frac{2b}{1-b}\gamma + \frac{b+b^2}{(1-b)^2} }. \\
	&= \frac{2\D}{3}  \frac{ \gamma^3 + \frac{3b}{1-b}\gamma^2 + \frac{3(b^2+b)}{(1-b)^2}\gamma  + b \frac{1+4b+b^2}{(1-b)^3}}{ \gamma^2 + \frac{2b}{1-b}\gamma + \frac{b+b^2}{(1-b)^2}  }. \label{eqn:cost2dim}
\end{align}

% for mathematica
% numerator =  r^3 +  3*b/(1-b)*r^2 +  3*(b^2+b)/ ( (1-b)^2) * r  + b* (1+4b+b^2)/((1-b)^3)

% denominator = r^2 +  2*b/(1-b)*r  +  (b+b^2)/ ( (1-b)^2 )

% coeff = 2*D/3

Therefore, in the two-dimensional setting, the optimal $\gamma^*$ is
\begin{align}
 	\gamma^* = \mathop{\arg \min}_{\gamma \in [0,1]} \frac{ \gamma^3 + \frac{3b}{1-b}\gamma^2 + \frac{3(b^2+b)}{(1-b)^2}\gamma  + b \frac{1+4b+b^2}{(1-b)^3}}{ \gamma^2 + \frac{2b}{1-b}\gamma + \frac{b+b^2}{(1-b)^2}  }.
 \end{align}

\begin{figure}[t]
\centering
\includegraphics[scale=0.4]{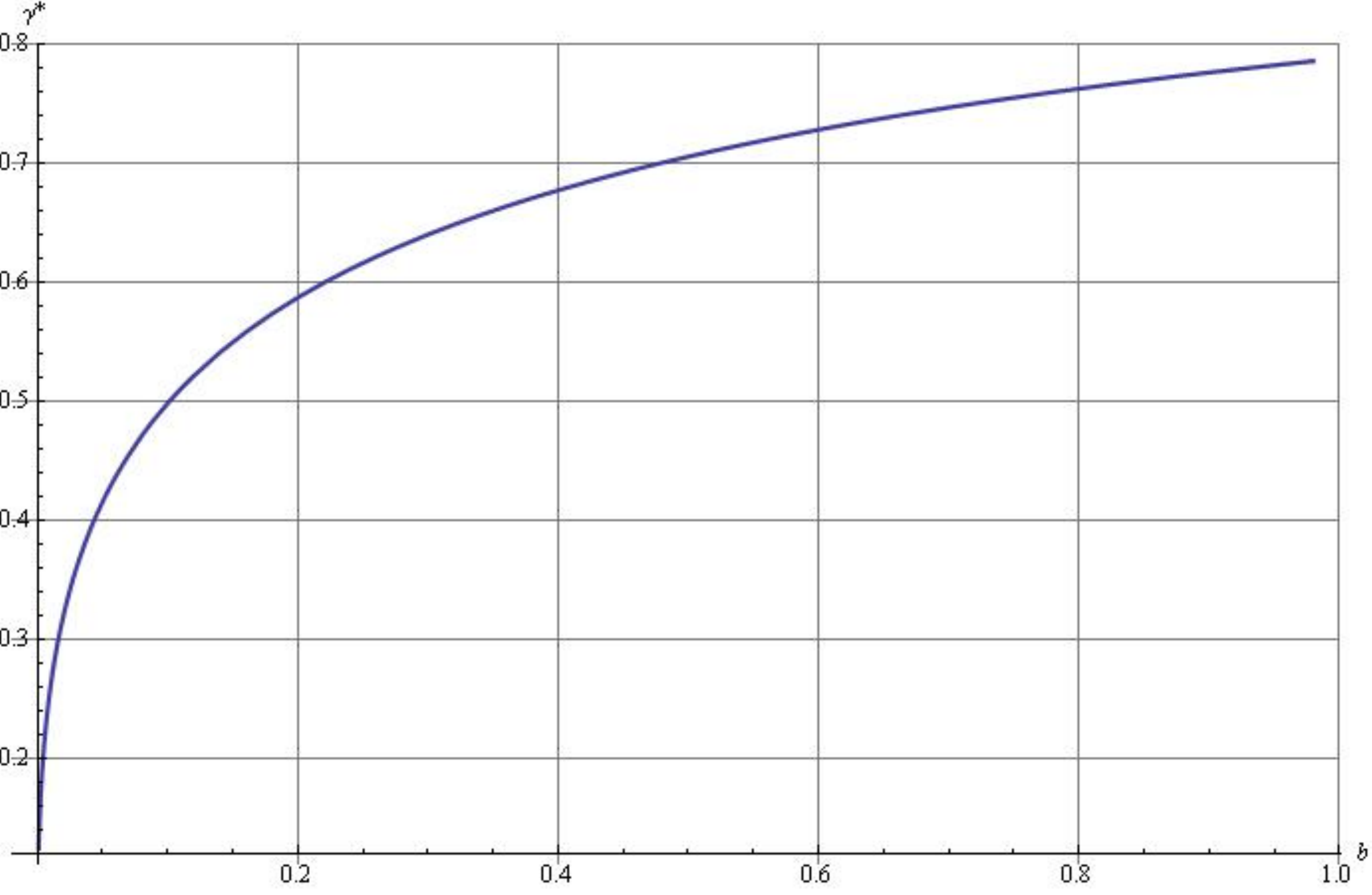}
\caption{The optimal $\gamma^*$ as a function of $b$}
\label{fig:gammastar2}
\end{figure}

 By setting the derivative of \eqref{eqn:cost2dim} to be zero, we use the software  \emph{Mathematica} to get a closed-form expression for $\gamma^*$, which is too complicated to show here. We plot $\gamma^*$ as a function of $b$ in Figure \ref{fig:gammastar2}.

% \begin{verbatim}
% 	\begin{align}
% 	\gamma^*=\frac{b}{b-1}+\frac{1}{2} \sqrt{\frac{\sqrt[3]{108 b^4+216 b^3+108 b^2+\sqrt{11664 b^8+46656 b^7-116640 b^6+46656 b^5+11664 b^4}}}{3 \sqrt[3]{2} \sqrt[3]{b^6-6 b^5+15 b^4-20 b^3+15 b^2-6 b+1}}-\frac{2 b^2}{(b-1)^2}+\frac{2 b^2}{b^2-2 b+1}+\frac{12 \sqrt[3]{2} b^2 \sqrt[3]{b^6-6 b^5+15 b^4-20 b^3+15 b^2-6 b+1}}{(b-1)^4 \sqrt[3]{108 b^4+216 b^3+108 b^2+\sqrt{11664 b^8+46656 b^7-116640 b^6+46656 b^5+11664 b^4}}}}+\frac{1}{2} \sqrt{\frac{2 b^2}{(b-1)^2}-\frac{2 b^2}{b^2-2 b+1}-\frac{\sqrt[3]{108 b^4+216 b^3+108 b^2+\sqrt{11664 b^8+46656 b^7-116640 b^6+46656 b^5+11664 b^4}}}{3 \sqrt[3]{2} \sqrt[3]{b^6-6 b^5+15 b^4-20 b^3+15 b^2-6 b+1}}-\frac{12 \sqrt[3]{2} b^2 \sqrt[3]{b^6-6 b^5+15 b^4-20 b^3+15 b^2-6 b+1}}{(b-1)^4 \sqrt[3]{108 b^4+216 b^3+108 b^2+\sqrt{11664 b^8+46656 b^7-116640 b^6+46656 b^5+11664 b^4}}}+\frac{\frac{16 \left(2 b^2+b\right)}{(b-1)^2}-\frac{32 b^3}{(b-1)^3}}{4 \sqrt{\frac{\sqrt[3]{108 b^4+216 b^3+108 b^2+\sqrt{11664 b^8+46656 b^7-116640 b^6+46656 b^5+11664 b^4}}}{3 \sqrt[3]{2} \sqrt[3]{b^6-6 b^5+15 b^4-20 b^3+15 b^2-6 b+1}}-\frac{2 b^2}{(b-1)^2}+\frac{2 b^2}{b^2-2 b+1}+\frac{12 \sqrt[3]{2} b^2 \sqrt[3]{b^6-6 b^5+15 b^4-20 b^3+15 b^2-6 b+1}}{(b-1)^4 \sqrt[3]{108 b^4+216 b^3+108 b^2+\sqrt{11664 b^8+46656 b^7-116640 b^6+46656 b^5+11664 b^4}}}}}}
% \end{align}
% \end{verbatim}

The optiaml cost $V^* = V(\p_{\gamma^*})$. We use Mathematica to analyze the asymptotic behavior of $V^*$ as $\e \to 0$ and $\e \to +\infty$. Indeed, we have
\begin{corollary}
In the high privacy regime,
\begin{align}
	V^* &= \frac{2\D}{\e} - \frac{\D \e^2}{36\sqrt{3}} + O(\e^3), \e \to 0,
\end{align}
and in the low privacy regime,
\begin{align}
	V^* &= \sqrt[3]{2}\D e^{-\frac{\e}{3}} + \frac{\D e^{-\frac{2\e}{3}} }{\sqrt[3]{2}} + o(e^{-\frac{2\e}{3}}), \e \to +\infty.
\end{align}
\end{corollary}

The Laplacian mechanism adds independent Laplacian noise to each component of the query output, and the cost is $\frac{2\D}{\e}$. Therefore, in the high privacy regime, the gap between optimal cost and the cost achieved by Laplacian mechanism goes to zero, as $\e \to 0$, and we conclude Laplacian mechanism is approximately optimal in the high privacy regime. However, in the low privacy regime (as $\e \to +\infty$), the optimal cost is  proportional to $e^{-\frac{\e}{3}}$, while the cost of Laplacian mechanism is proportional to $\frac{1}{\e}$. We conclude the gap is significant in the low privacy regime.

It is natural to compare the performance of the optimal multi-dimensional staircase mechanism and the composite single-dimensional staircase mechanism which adds independent staircase noise to each component of the query output. If independent staircase noise is added to each component of query output, to satisfy the $\e$-differential privacy constraint, the parameter of the  staircase noise  is $\frac{\e}{2}$ instead of $\e$, and thus the total cost will be proportional to $e^{-\frac{\e}{4}}$, which is worse than the optimal cost $\Theta(e^{-\frac{\e}{3}})$.

\section{Proof of Theorem \ref{thm:main}} \label{sec:proof}
In this section, we give a detailed  proof of Theorem \ref{thm:main}.

\subsection{Outline of Proof}
 The key idea of the proof is to use a sequence of probability distributions with piecewise constant probability density functions to approximate any probability distribution satisfying the differential privacy constraint \eqref{eqn:dpconstraintfinal}.  The proof consists of 4 steps in total, and in each step we narrow down the set of  probability distributions where the optimal probability distribution should lie in:

 \begin{itemize}
 	\item Step 1 proves that we only need to consider  probability distributions which have symmetric piecewise constant probability density functions.

 	\item Step 2 proves that we only need to consider those symmetric piecewise constant probability density functions which are monotonically decreasing.

 	\item Step 3 proves that optimal probability density function should periodically decay.

 	\item Step 4 proves that the optimal probability density function is staircase-shaped in the multidimensional setting, and it concludes the proof of Theorem \ref{thm:main}.
 \end{itemize}

%%%%%%%%%%%%%%%%%%%%% first step, symmetric %%%%%
\subsection{Step 1}

Given $\p \in \sP$, define
\begin{align}
	V(\p) \triangleq \int \int \dots \int_{\R^d} \loss(\bx) \p (d x_1 d x_2 \dots d x_d).
\end{align}

Define
\begin{align}
 	V^* \triangleq \inf_{\p \in \sP}  V(\p). \label{def:Vstar}
 \end{align}
Our goal is to prove that $V^* =  \inf\limits_{\gamma \in [0,1]}  \int \int \dots \int_{\R^d} \loss(\bx) f_\gamma(\bx)  d x_1 d x_2 \dots d x_d $.

If $V^* = + \infty$, then due to the definition of $V^*$, we have
\begin{align}
	 \inf_{\gamma \in [0,1]}  \int \int \dots \int_{\R^d} \loss(\bx)     f_{\gamma}(\bx)  d x_1 d x_2 \dots d x_d  \ge V^* = + \infty,
\end{align}
and thus $\inf_{\gamma \in [0,1]}  \int \int \dots \int_{\R^d} \loss(\bx)     f_{\gamma}(\bx)  d x_1 d x_2 \dots d x_d  = V^* = + \infty$. So we only need to consider the case $V^* < +\infty$, i.e., $V^*$ is finite. Therefore, in the rest of the proof, we assume $V^*$ is finite.

First we show that given any probability measure $\p \in \sP$, we can use a sequence of probability measures  with multidimensionally piecewise constant probability density functions to approximate $\p$.

% with $\lim_{i \to +\infty} V(\p_i) = V(\p)$.

Given $i \in \N$ and  $k \in \N$, define
\begin{align}
	A_i(k) =  \{ \bx \in \R^d | k\frac{\D}{i} \le \|\bx\|_1 < (k+1)\frac{\D}{i}  \} \subset \R^d.
\end{align}

It is easy to calculate the volumn of $A_i(k)$, which is
\begin{align}
	\text{Vol} (A_i(k)) = \frac{2^d}{d!} \left( (k+1)^d - k^d \right) \frac{\D^d}{i^d}.
\end{align}.

\begin{lemma}\label{lem:approx}
	Given $\p \in \sP$ with $V(\p) < + \infty$, any positive integer $i \in \N$, define  $\p_i$  as the probability distribution with probability density function $f_i(\bx)$ defined as
	\begin{align}
		f_i(\bx) = a_i(k) \triangleq \frac{ \p( A_i (k) ) }{\text{Vol}(A_i(k))}  & \bx \in A_i(k) \;  \text{for} \; k \in \N .	 \label{eqn:aik}
	\end{align}

	Then $\p_i \in \sP$ and
	\begin{align}
		\lim_{i \to +\infty} V(\p_i) = V(\p).
	\end{align}
	
\end{lemma}

We conjecture that Lemma \ref{lem:approx} holds for arbitrary dimension $d$, and prove it for the case $d=2$.

Before proving Lemma \ref{lem:approx} for $d=2$, we prove an auxiliary Lemma which shows that for probability mass function over $\Z^2$ satisfying $\e$-differential privacy constraint, we can construct a new probability mass function by averaging the old probability mass function over each $\ell^1$ ball and the new probability mass function still satisfies the $\e$-differential privacy constraint.

%%%%%%%%%%%%%%%%%%%%%%%%%%%%% Averaging Lemma %%%%%%%%%

% for dimension 2, discrete case, probability mass function.

\begin{lemma}\label{lem:averagediscrete}
	For any given probability mass function $\p$ defined over the set $\Z^2$ satisfying that
	\begin{align}
		\p(i_1,j_1) \le e^{\e}\p(i_2,j_2), \forall |i_1 - i_2| + |j_1 - j_2| \le \D, \label{eqn:dpdiscrete5}
	\end{align}
	define the probability mass function $\tilde{\p}$ via
	\begin{align}
		\tilde{\p}(i,j) =
		\begin{cases}
			\p(0,0) & (i,j) = (0,0) \\
			p_{|i| + |j|} & (i,j) \neq (0,0)
		\end{cases}
	\end{align}
	where $ p_k \triangleq \frac{\sum_{(i',j') \in \Z^2 : |i'| + |j'| = k } \p(i',j')}{4 k}, \forall k \ge 1$.

	Then $\tilde{\p}$ is also a probability mass function satisfying the differential privacy constraint, i.e.,
	\begin{align}
		\tilde{\p}(i_1,j_1) \le e^{\e}\tilde{\p}(i_2,j_2), \forall |i_1 - i_2| + |j_1 - j_2| \le \D. \label{eqn:2dimdp}
	\end{align}
\end{lemma}

\begin{IEEEproof}
	Due to the way how we define $\tilde{\p}$, we have
	\begin{align}
		\sum_{(i,j) \in \Z^2} \tilde{\p}(i,j) = \sum_{(i,j) \in \Z^2} \p(i,j) = 1,
	\end{align}
	and thus $\tilde{\p}$ is a valid probability mass function defined over $\Z^2$.

	Next we prove that $\tilde{\p}$ satisfies \eqref{eqn:2dimdp}. To simplify notation, define $p_0 \triangleq \p(0,0)$. Then we only need to prove that for any $k_1,k_2 \in \N$ such that $|k_1 - k_2| \le \D$, we have
	\begin{align}
		p_{k_1} \le e^{\e} p_{k_2}.
	\end{align}

	Due to the symmetry property, without loss of generality, we can assume $k_1 < k_2$.

	The easiest case is $k_1 = 0$. When $k_1 = 0$, we have $k_2 \le \D$ and
	\begin{align}
		\p(0,0) \le e^{\e} \p(i,j), \forall |i|+|j| = k_2. \label{eqn:sumpt1}
	\end{align}

	The number of distinct pairs $(i,j)$ satisfying $|i| + |j| = k $ is $4k$ for $k \ge 1$. Sum up all inequalities in \eqref{eqn:sumpt1}, and we get
	\begin{align}
		& 4 k_2 \p(0,0) \le e^{\e} \sum_{(i,j) \in \Z^2: |i|+|j| = k_2 } \p(i,j) \\
		\Leftrightarrow  & \p(0,0) \le e^{\e} \frac{\sum_{(i,j) \in \Z^2: |i|+|j| = k_2 } \p(i,j)}{4 k_2} \\
		\Leftrightarrow  & p_0 \le e^{\e} p_{k_2}.
	\end{align}

	For general $0<k_1<k_2$, let $\dd \triangleq k_2 - k_1 \le \D$. Define $B_k$ via
	\begin{align}
	 	B_k \triangleq \{(i,j) \in \Z^2 | |i| + |j| = k\}, \forall k \in \N.
	 \end{align}

	Then the differential privacy constraint \eqref{eqn:dpdiscrete5} implies that
	\begin{align}
		\p(i_1, j_1) \le e^{\e} \p(i_2,j_2), \forall (i_1,j_1) \in B_{k_1}, (i_2,j_2) \in B_{k_2}, |i_1 - i_2| + |j_1 - j_2| = \dd. \label{eqn:rect1}
	\end{align}

	The set of points in $B_k$ forms a rectangle, which has $4$ corner points and $4(k-1)$ interior points on the edges. For each corner point in $B_{k_1}$, which appears in the left side of \eqref{eqn:rect1}, there are $(2\dd + 1)$ points in $B_{k_2}$ close to it with an $\ell^1$ distance of $\dd$. And for each interior point in $B_{k_1}$, there are $(\dd+1)$ points in $B_{k_2}$ close to it with an $\ell^1$ distance of $\dd$. Therefore, there are in total $4(2\dd+1) + 4(k_1 - 1)(\dd+1)$ distinct inequalities in \eqref{eqn:rect1}.

	If we can find certain nonnegative coefficients such that  multiplying each inequality in \eqref{eqn:rect1} by these nonnegative coefficients and summing them up  gives us
	\begin{align}
		\frac{\sum_{(i',j') \in \Z^2 : |i'| + |j'| = k_1 } \p(i',j')}{4 k_1} \le e^{\e} \frac{\sum_{(i',j') \in \Z^2 : |i'| + |j'| = k_2 } \p(i',j')}{4 k_2},
	\end{align}
	then \eqref{eqn:2dimdp} holds. Therefore, our goal is to find the ``right'' coefficients associated with each inequality in \eqref{eqn:rect1}. We formulate it as a matrix filling-in problem in which we need to choose nonnegative coefficients for certain entries in a matrix such that the sum of each row is $\frac{k_1 + \dd}{k_1}$, and the sum of each column is 1.

	More precisely, label the $4 k_1$ points in $B_{k_1}$ by $\{I_1,I_2,I_3, \dots, I_{4k_1}\}$, where we label the topmost point by $1$ and sequentially label other points clockwise. Similarly, we label the $4 k_2$ points in $B_{k_2}$ by $\{O_1,O_2,O_3, \dots, O_{4k_2}\}$, where we label the topmost point by $1$ and sequentially label other points clockwise.

	Consider the following $4k_1$ by $4k_2$ matrix $M$, where each row corresponds to the point in $B_{k_1}$ and each column corresponds to the point in $B_{k_2}$, and the entry $M_{ij}$ in the $i$th row and $j$th column is the coefficient corresponds to inequality involved with the points $I_i$ and $O_j$. If there is no inequality associated with the points $I_i$ and $O_j$, then $M_{ij} = 0$.

	In the case $k_1 = 2$ and $\dd = 3$, the zeros/nonzeros pattern of $M$ has the following form:
\begin{align}
	\left(
  \begin{array}{cccccccccccccccc}
    x & x & x & 0 & 0 & 0 & 0 & 0 & 0 & 0 & 0 & 0 & 0 & 0 & x & x \\
    0 & x & x & x & 0 & 0 & 0 & 0 & 0 & 0 & 0 & 0 & 0 & 0 & 0 & 0 \\
    0 & 0 & x & x & x & x & x & 0 & 0 & 0 & 0 & 0 & 0 & 0 & 0 & 0 \\
    0 & 0 & 0 & 0 & 0 & x & x & x & 0 & 0 & 0 & 0 & 0 & 0 & 0 & 0 \\
    0 & 0 & 0 & 0 & 0 & 0 & x & x & x & x & x & 0 & 0 & 0 & 0 & 0 \\
    0 & 0 & 0 & 0 & 0 & 0 & 0 & 0 & 0 & x & x & x & 0 & 0 & 0 & 0 \\
    0 & 0 & 0 & 0 & 0 & 0 & 0 & 0 & 0 & 0 & x & x & x & x & x & 0 \\
    0 & 0 & 0 & 0 & 0 & 0 & 0 & 0 & 0 & 0 & 0 & 0 & 0 & x & x & x \\
  \end{array}
\right),
\end{align}
where $x$ denotes an entry which can take any nonnegative coefficient.

For general $k_1$ and $k_2$, the pattern of $M$ is that the first, $(k_1 + 1)$th, $(2 k_1 + 1)$th and $(3k_1 + 1)$th rows can have $2\dd + 1$ nonzero entries, and all other rows can have $\dd+1$ nonzero entries.

	We want to show that
	\begin{align}
		\frac{\sum_{(i',j') \in \Z^2 : |i'| + |j'| = k_1 } \p(i',j')}{4 k_1} \le e^{\e} \frac{\sum_{(i',j') \in \Z^2 : |i'| + |j'| = k_2 } \p(i',j')}{4 k_2},
	\end{align}
	or equivalently,
	\begin{align}
		 (1 + \frac{\dd}{k_1}) \sum_{(i',j') \in \Z^2 : |i'| + |j'| = k_1 } \p(i',j')  \le e^{\e} \ \sum_{(i',j') \in \Z^2 : |i'| + |j'| = k_2 } \p(i',j').
	\end{align}

Therefore, our goal is to find nonnegative coefficients to substitute each $x$ in the matrix such that the sum of each column is 1 and the sume of each column is $(1 + \frac{\dd}{k_1})$.
We will give explicit formulas on how to choose the coefficients.

The case $k_1 = 1$ is trivial. Indeed, one can set all diagonal entries to be 1, and set all other nonzero entries to be $\frac{1}{2}$. Therefore, we can assume $k_1 > 1$.

Consider two different cases: $k_1 \le \dd$ and $k_1 \ge \dd + 1$.

We first consider the case $k_1 \le \dd$. Due to the periodic patterns in $M$, we only need to consider rows from $1$ to $k_1 + 1$. Set all entries to be zero except that we set
\begin{align}
  M_{11} &= M_{22} = \cdots = M_{k_1 k_1} = 1, \\
  M_{2,\dd+2} &= M_{3,\dd+3} = \cdots = M_{k_1+1,k_1+\dd + 1 } = 1 \\
  M_{1,j} &= \frac{\dd}{2k_1 (\dd-k_1+1)},  j \in [k_1+1,\dd+1] \cup   [ 4k_1 - \dd + 1  , 4k_1]\\
  M_{k_1+1,j} &= \frac{\dd}{2k_1 (\dd-k_1+1)},  j \in [k_1+1,\dd+1] \cup [2k_1+1+\dd, k_1 + 1 + 2\dd] \\
  M_{i,j} &= \frac{1 - \frac{\dd}{k_1 (\dd-k_1 + 1)}}{k_1-1}.
 \end{align}

It is straightforward to verify that the above matrix $M$ satisfies the properties that the sum of each column is $1$ and the sum of each row is $(1 + \frac{\dd}{k_1})$. Therefore, we have
\begin{align}
	p_{k_1} \le e^{\e} p_{k_2}, \forall 0<k_1<k_2, k_1 \le k_2 - k_1 \le \D.
\end{align}

Next we solve the case $k_1 \ge \dd + 1$. Again due to the periodic patterns in $M$, we only need to consider the nonzero entries in rows from $1$ to $k_1 + 1$. We use the following procedures to construct $M$:
\begin{enumerate}
	\item For the first row, set $M_{11} = 1$ and set all other $2\dd$ nonzero entries to be $\frac{1}{2k_1}$.
	\item For the second row, $M_{22}$ is uniquely determined to be $1 - \frac{1}{2k_1}$. Set the next $\dd - 1$ nonzero entries in the second row to be $\frac{1}{k_1}$, i.e., $M_{2j} = \frac{1}{k_1}$ for $j \in [3, \dd+1]$. The last nonzero entry $M_{2,\dd+2}$ is uniquely determined to be
	\begin{align}
	 	(1+\frac{\dd}{k_1}) - (1-\frac{1}{2k_1}) - \frac{\dd-1}{k_1} = \frac{3}{2k_1}.
	 \end{align}
	\item For the third row, the first nonzero entry $M_{33}$ is  uniquely determined to be $1 - \frac{1}{2k_1} - \frac{1}{k_1} = 1 - \frac{3}{2k_1}$. Set the next $\dd - 1$ nonzero entries to be $\frac{1}{k_1}$, i.e., $M_{3j} = \frac{1}{k_1}$ for $j \in [4, \dd+2]$. The last nonzero entry $M_{3,\dd+3}$ is uniquely determined to be
	\begin{align}
	 	(1+\frac{\dd}{k_1}) - (1-\frac{3}{2k_1}) - \frac{\dd-1}{k_1} = \frac{5}{2k_1}.
	 \end{align}

	 \item In general, for the $i$th row ($i \in [2, k_1 -1]$), the first nonzero entry $M_{ii}$ is set to be   $M_{ii} = 1 - \frac{2i-3}{2k_1}$, and the next $\dd-1$ nonzero entries are $\frac{1}{k_1}$, and the last nonzero entry $M_{i,i+\dd} = \frac{2i-1}{2k_1}$.
	 \item For $(k_1+1)$th row, by symmetry, we set $M_{k_1+1,k_1+1} = 1$ and set other $2\dd$ nonzero entries to be $\frac{1}{2k_1}$.
	 \item The nonzero entries in the $k_1$th row are uniquely determined. Indeed, we have
	 \begin{align}
	 	M_{k_1, k_1} &= 1 - \frac{2k_1 - 3}{2k_1}, \\
	 	M_{k_1, k_1 + \dd} &= 1 - \frac{1}{2k_1}, \\
	 	M_{k_1, k_1 + j} &= \frac{1}{k_1}, j \in [2,\dd-1].
	 \end{align}
\end{enumerate}

It is straightforward to verify that each entry in $M$ is nonnegative and $M$ satisfies the properties that the sum of each column is $1$ and the sum of each row is $(1 + \frac{\dd}{k_1})$. Therefore, we have
\begin{align}
	p_{k_1} \le e^{\e} p_{k_2}, \forall 0<k_1<k_2, k_1 \ge \dd + 1 = k_2 - k_1 + 1 .
\end{align}

Therefore, for all $k_1,k_2 \in \N$ such that $|k_2-k_1| \le \D$, we have
\begin{align}
	p_{k_1} \le e^{\e} p_{k_2}.
\end{align}

This completes the proof of Lemma \ref{lem:averagediscrete}.

\end{IEEEproof}

%%%%%%%%%%%%%%%%%%%%%%%%%%%%  end of averaging over discrete lattice

\begin{IEEEproof}[Proof of Lemma \ref{lem:approx}]

	First we prove that $\p_i \in \sP$, i.e., $\p_i$ satisfies the differential privacy constraint \eqref{eqn:dpconstraintfinal}.

	By the definition of  $f_i(\bx)$,  $f_i(\bx)$ is a nonnegative function, and
	\begin{align}
		& \int \int \dots \int_{\R^d} f_i (\bx) d x_1 d x_2 \dots d x_d \\
		=& \sum_{k=0}^{+\infty} a_i(k) \text{Vol}(A_i(k))  \\
		=&  \sum_{k=0}^{+\infty} \p( A_i (k) )  \\
		=&  \p(\R^d ) = 1.
	\end{align}
	So $\p_i$ is a valid probability distribution.

Next we show that $f_i(\bx)$ satisfies the differential privacy constraint.
For fixed $i$, on the $x_1 - x_2$ plane, we can use the lines $x_2 = x_1 + \frac{k}{i}\D$ and $x_2 = -x_1 + \frac{k}{i}\D$ for all $k \in \Z$ to divide each $A_i(k)$ into distinct squares with the same size (each $A_i(k)$ will be divided into $8k+4$ squares). By taking the average of the probability density function over each square, we reduce the probability density function to a discrete probability mass function over $\Z^2$ satisfying $\e$-differential privacy constraint. Then apply Lemma \ref{lem:averagediscrete}, and we have
\begin{align}
	a_i(k_1) \le e^{\e} a_i(k_2), \forall k_1,k_2 \in \N \; \text{with} \; |k_1 - k_2| \le i.
\end{align}

Given $\bx, \by \in \R^d$ such that $\|\bx-\by\|_1 \le \D$, let $k_1, k_2$ be the integers such that
\begin{align}
	\bx &\in A_i (k_1), \\
	\by &\in A_i (k_2).
\end{align}
Then $|k_1 - k_2| \le i$. Therefore,
\begin{align}
	f_i (\bx) \le e^{\e} f_i(\by),
\end{align}
which implies that the probability distribution $\p_i$ satisfies the differential privacy constraint \eqref{eqn:dpconstraintfinal}.

Therefore, for any  integer $i \ge 1$, $\p_i \in \sP$.

Next we show that
	\begin{align}
		\lim_{i \to +\infty} V(\p_i) = V(\p).
	\end{align}

Given $\delta >0$, since $V(\p)$ is finite, there exists $T^* = m \D >1$ for some $m \in \N$ such that
\begin{align}
	\int \int \dots \int_{ \{ \bx \in \R^d | \|\bx\|_1 \ge T^*\} } \loss(\bx)  \p(d x_1 d x_2 \dots d x_d) < \frac{\delta}{2}.
\end{align}

For each $A_i(k)$  we have
\begin{align}
	\int \int \dots \int_{A_i(k) } \loss(\bx)  \p_i(d x_1 d x_2 \dots d x_d) &= \int \int \dots \int_{A_i(k) } \|\bx\|_1  \p_i(d x_1 d x_2 \dots d x_d) \\
	&\le \p_i(A_i(k))  (k+1)\frac{\D}{i}  \\
	&= \p(A_i(k))  (k+1)\frac{\D}{i}   \\
	&\le 2 \p(A_i(k))  k\frac{\D}{i}   \\
	&\le 2 \int \int \dots \int_{A_i(k) } \loss(\bx)  \p(d x_1 d x_2 \dots d x_d).
\end{align}

Therefore,
\begin{align}
\int \int \dots \int_{\{ \bx \in \R^d | \|\bx\|_1 \ge T^*\} } \loss(\bx)  \p_i(d x_1 d x_2 \dots d x_d) &\le 2 \int \int \dots \int_{\{ \bx \in \R^d | \|\bx\|_1 \ge T^*\} } \loss(\bx)  \p(d x_1 d x_2 \dots d x_d) \\
&\le 2 \frac{\delta}{2} = \delta.
\end{align}

 $\loss(\bx)$ is a bounded function when $\|\bx \|_1 \le T^*$, and thus by the definition of  Riemann$\text{-}$Stieltjes integral, we have
\begin{align}
	\lim_{i \to \infty} \int \int \dots \int_{\{ \bx \in \R^d | \|\bx\|_1 < T^*\}  } \loss(\bx)  \p_i(d x_1 d x_2 \dots d x_d) = \int \int \dots \int_{\{ \bx \in \R^d | \|\bx\|_1 < T^*\} } \loss(\bx)  \p(d x_1 d x_2 \dots d x_d).
\end{align}

So  there exists a sufficiently large integer  $i^*$ such that for all $i \ge i^*$
\begin{align}
	\left | \int \int \dots \int_{\{ \bx \in \R^d | \|\bx\|_1 < T^*\} } \loss(\bx)  \p_i(d x_1 d x_2 \dots d x_d)  - \int \int \dots \int_{\{ \bx \in \R^d | \|\bx\|_1 < T^*\}  } \loss(\bx)  \p(d x_1 d x_2 \dots d x_d) \right |  \le \delta.
\end{align}

To simplify notation, we use $d \bx $ to denote $d x_1 d x_2 \dots d x_d$.

Hence, for all $i \ge i^*$
\begin{align}
	&\; |V(\p_i) - V(\p)| \\
	=&\; \left |  \int_{\R^d } \loss(\bx)  \p_i(d \bx )  -   \int_{\R^d } \loss(\bx)  \p(d \bx )  \right | \\
	=&\;  \left | \int_{\{ \bx \in \R^d | \|\bx\|_1 < T^*\} } \loss(\bx)  \p_i(d \bx )  -   \int_{\{ \bx \in \R^d | \|\bx\|_1 < T^*\} } \loss(\bx)  \p(d \bx )
	+    \int_{\{ \bx \in \R^d | \|\bx\|_1 \ge T^*\}} \loss(\bx)  \p_i(d \bx )  -   \int_{\{ \bx \in \R^d | \|\bx\|_1 \ge T^*\} } \loss(\bx)  \p(d \bx ) \right | \\
	\le &\;  \left |  \int_{\{ \bx \in \R^d | \|\bx\|_1 < T^*\}  } \loss(\bx)  \p_i(d \bx )  -   \int_{\{ \bx \in \R^d | \|\bx\|_1 < T^*\} } \loss(\bx)  \p(d \bx )  \right |
	+    \int_{ \{ \bx \in \R^d | \|\bx\|_1 \ge T^*\} } \loss(\bx)  \p_i(d \bx )  +    \int_{\{ \bx \in \R^d | \|\bx\|_1 \ge T^*\}} \loss(\bx)  \p(d \bx) \\
	\le &\;  (\delta + \delta + \frac{\delta}{2}) \\
	\le &\;  \frac{5}{2} \delta.
\end{align}
% \begin{align}
% 	&\; |V(\p_i) - V(\p)| \\
% 	=&\; \left | \int \int \dots \int_{\R^d } \loss(\bx)  \p_i(d x_1 d x_2 \dots d x_d)  - \int \int \dots \int_{\R^d } \loss(\bx)  \p(d x_1 d x_2 \dots d x_d)  \right | \\
% 	=&\;  \left |\int \int \dots \int_{\{ \bx \in \R^d | \|\bx\|_1 < T^*\} } \loss(\bx)  \p_i(d x_1 d x_2 \dots d x_d)  - \int \int \dots \int_{\{ \bx \in \R^d | \|\bx\|_1 < T^*\} } \loss(\bx)  \p(d x_1 d x_2 \dots d x_d)
% 	+  \int \int \dots \int_{\{ \bx \in \R^d | \|\bx\|_1 \ge T^*\}} \loss(\bx)  \p_i(d x_1 d x_2 \dots d x_d)  - \int \int \dots \int_{\{ \bx \in \R^d | \|\bx\|_1 \ge T^*\} } \loss(\bx)  \p(d x_1 d x_2 \dots d x_d) \right | \\
% 	\le &\;  \left |\int \int \dots \int_{\{ \bx \in \R^d | \|\bx\|_1 < T^*\}  } \loss(\bx)  \p_i(d x_1 d x_2 \dots d x_d)  - \int \int \dots \int_{\{ \bx \in \R^d | \|\bx\|_1 < T^*\} } \loss(\bx)  \p(d x_1 d x_2 \dots d x_d)  \right |
% 	+  \int \int \dots \int_{ \{ \bx \in \R^d | \|\bx\|_1 \ge T^*\} } \loss(\bx)  \p_i(d x_1 d x_2 \dots d x_d)  +  \int \int \dots \int_{\{ \bx \in \R^d | \|\bx\|_1 \ge T^*\}} \loss(\bx)  \p(d x_1 d x_2 \dots d x_d) \\
% 	\le &\;  (\delta + \delta + \frac{\delta}{2}) \\
% 	\le &\;  \frac{5}{2} \delta.
% \end{align}

Therefore,
\begin{align}
	\lim_{i \to +\infty}  V(\p_i) = V(\p).
\end{align}

\end{IEEEproof}

Define $\sPsymi  \triangleq \{\p_i | \p \in \sP \}$ for $ i \ge 1$, i.e., $\sPsymi$ is the set of probability distributions satisfying differential privacy constraint \eqref{eqn:dpconstraintfinal} and having symmetric piecewise constant (over $A_i(k) \; \forall k \in \N$) probability density functions.

Due to Lemma \ref{lem:approx},
\begin{lemma}\label{lem:piecewiseconstant}
	\begin{align}
	V^*  = \inf_{\p \in \cup_{i=1}^{\infty} \sPsymi}  V(\p).
\end{align}
\end{lemma}

Therefore, to characterize $V^*$, we only need to study probability distributions with symmetric and piecewise constant probability density functions.

%%%%%%%%%%%%%%%%%%%%%%%%%%%%% step 2  %%%%%%%%%%%%%%%%%%%%%
\subsection{Step 2}

Given $\p \in \psym$, we call $\{a_i(0),a_i(1),a_i(2),\dots\}$ the density sequence of $\p_i \in \sPsymi$, where $a_i(k)$ is defined in \eqref{eqn:aik} $\forall k \in \N$.

Next we show that indeed we only need to consider those probability distributions with symmetric piecewise constant probability density functions the density sequences of which are \emph{monotonically decreasing}.

Define
\begin{align}
	\pa \triangleq \{ \p | \p \in \sPsymi, \; \text{and} \; \text{the density sequence of}\; \p \; \text{is  monotonically decreasing} \}.
\end{align}
Then
\begin{lemma}\label{lem:monotone}
	\begin{align}
	V^* = \inf_{\p \in \cup_{i=1}^{\infty} \pa}   V(\p)   .
\end{align}
\end{lemma}

\begin{IEEEproof}
We first show that among  $\sPsymi$, to mininize the cost we only need to consider these probability distributions with density sequences $\{a_0,a_1,a_2,\dots \}$ satisfying that $a_0 \ge a_1$. Indeed, given $\p_a \in \sPsymi$ with   density sequence $\{a_0,a_1,a_2,\dots \}$ such that $a_0 < a_1$,  there exists $\p_b \in \sPsymi$ with density sequence $\{b_0,b_1,b_2,\dots\}$ such that $b_0 \ge b_1$ and
\begin{align}
	V(\p_b) \le V(\p_a).
\end{align}

Consider the probability distribution $\p_b \in \sPsymi$ with density sequence $\{b_0,b_1,b_2,,\dots\}$ defined as
\begin{align}
	b_0 &= (1+\delta) a_0, \\
	b_k &= (1 - \delta') a_k,  \forall k \ge 1,
\end{align}
where we choose $\delta >0 $ and $0 <\delta' < 1 $ such that
\begin{align}
	b_0 &= b_1, \label{eqn:constr1} \\
	\sum_{k=0}^{+\infty} b_k \text{Vol}(A_i(k)) &= \sum_{k=0}^{+\infty} a_k \text{Vol}(A_i(k)) = 1. \label{eqn:constr2}.
\end{align}

Equation \eqref{eqn:constr2} makes $\p_b$ be a valid probability distribution. One can easily solve \eqref{eqn:constr1}  and \eqref{eqn:constr2}, and write down the explicit expression for $\delta, \delta'$. The density sequence $\{b_0,b_1,b_2,\dots\}$ satisfies $b_0 \ge b_1$ (indeed, we have $b_0 = b_1$), and it is easy to check it satisfies the differential privacy constraint, i.e.,
\begin{align}
	b_{k_1} \le e^{\e} b_{k_2}, \forall k_1,k_2 \in \N \; \text{with} \; |k_1 - k_2| \le i.
\end{align}

Note that $\cost(\|\bx\|_1)$ is a monotonically increasing function of $\|\bx\|_1$, and compared to $\p_a$, $\p_b$ moves some probability of $\pa$ from the (higher cost) area $\{\bx | \|bx\| \ge \frac{\D}{i}\}$  to the (lower cost) area  $\{\bx | \|bx\| \le \frac{\D}{i}\}$, and thus we have
\begin{align}
	V(\p_b) \le V(\p_a).
\end{align}

Therefore, among  $\sPsymi$, to mininize the cost we only need to consider these probability distributions with density sequences $\{a_1,a_2,a_3,\dots\}$ satisfying that $a_0 \ge a_1$.

Next we show that among $\sPsymi$ with density sequences $\{a_1,a_2,a_3,\dots\}$  satisfying $a_0 \ge a_1$, to mininize the cost we only need to consider these probability distributions with density sequences  also satisfying that $ a_1 \ge a_2$.

Given $\p_a \in \sPsymi$ with density sequence $\{a_1,a_2,a_3,\dots \}$ such that $a_0 \ge a_1$ and $a_1 < a_2$,  there exists $\p_b \in \sPsymi$ with density sequence $\{b_1,b_2,b_3,\dots\}$ such that $b_0 \ge b_1$ and
\begin{align}
	b_1 \ge b_2.
\end{align}

If $i \le 2$, we can construct $\p_b$ by scaling up $a_0,a_1$ and scale down $a_k$ for all $k \ge 2$. More precisely, define $\p_b$ with density sequence $\{b_0,b_1,b_2,\dots\}$ via
\begin{align}
	b_k &= (1+\delta)a_k, k \le 1, \\
	b_k &= (1-\delta')a_k, k \ge 2,
\end{align}
for some $\delta >0$ and $0 < \delta' < 1$ such that
\begin{align}
	b_2 &= b_1, \\
	\sum_{k=0}^{+\infty} b_k \text{Vol}(A_i(k)) &= \sum_{k=0}^{+\infty} a_k \text{Vol}(A_i(k)) = 1.
\end{align}
So we have $b_0 \ge b_1 \ge b_2$. It is easy to check that  $\p_b$ satisfies the differential privacy constraint, and $V(\p_b) \le V(\p_a)$ using the fact that $\cost(\|\bx\|_1)$ is a monotonically decreasing function in terms of $\|\bx\|_1$.

If $i \ge 3$, then without loss of generality we can assume $a_2 \le a_0$. Indeed, if $a_2 > a_0$, we can scale up $a_0,a_1$ and scale down $a_k$ for all $k \ge 2$ to make $a_2 = a_0$, and this operation will preserve the differential privacy constraint and decrease the cost. Note that, in this case we cannot use the same scaling operation to make $a_2 \le a_0$, because it is possible that after the scaling operation $\frac{a_0}{a_k} > e^{\e}$ for some $3 \le k \ge i$ violating the differential privacy constraint. Hence, we can assume $a_0 \ge a_2 > a_1$. Let $a_{k'}$ be the largest value in $\{a_3,\dots, a_{2+i} \}$. If $\frac{a_{k'}}{a_2} < e^{\e}$, we can scale up $a_1$ and scale down $a_2$ until $a_1 = a_2$ or $\frac{a_{k'}}{a_2} = e^{\e}$. It is easy to see this scaling operation will preserve differential privacy and decrease the cost. If after this scaling operation we have $a_2 = a_1$, then we are done. Suppose $a_1$ is still bigger than $a_2$. Then $a_2$ is the smallest element in $\{a_2,a_3,\dots,a_{2+i}\}$. Therefore, we have $\max_{2 \le k \le i} \frac{a_0}{a_k} = \frac{a_0}{a_2}$. Then we can scale up $a_0,a_1$ and scale down $a_k$ for $k \ge 2$ until $a_1 = a_2$. This operation will preserve the differential privacy constraint and decrease the cost. If we call the final probability distribution we obtained $\p_b$, we have $\p_b \in \sPsymi$, and the density sequence satisfying $b_0 \ge b_1 \ge b_2$ (indeed, $b_1 = b_2$), and $V(\p_b) \le V(\p_a)$.

By induction, we can show that among all probability distributions in $\sPsymi$, to mininize the cost we only need to consider probability distributions with monotonically decreasing density sequence.

Suppose among $\sPsymi$ to minimize the cost we only need to consider probability distribution  with density sequence $\{a_0,a_1,a_2,\dots\}$ satisfying $a_0 \ge a_1 \ge a_2 \ge \cdots \ge a_n$. Then we can show that among  $\sPsymi$ to minimize the cost we only need to consider probability distribution   with density sequence $\{a_0,a_1,a_2,\dots\}$ satisfying $a_0 \ge a_1 \ge a_2 \ge \cdots \ge a_n \ge a_{n+1}$.

Indeed, given $\p_a \in \sPsymi$ with density sequence $\{a_0,a_1,a_2,\dots\}$ satisfying $a_0 \ge a_1 \ge a_2 \ge \cdots \ge a_n$, we can construct $\p_b \in \sPsymi$ with density sequence $\{b_0,b_1,b_2,\dots\}$ satisfying
\begin{align}
	b_0 \ge b_1 \ge b_2 \ge \cdots \ge b_n \ge b_{n+1},
\end{align}
and
\begin{align}
	V(\p_b) \le V(\p_a).
\end{align}

If $a_{n+1} \le a_{n}$, then we can choose $\p_b = \p_a$.

Suppose $a_{n+1} > a_{n}$.  Without loss of generality, we can assume
\begin{align}
	a_{n+1} \le a_k, \; \text{for} \; k \le n+2-i. \label{eqn:less11}
\end{align}
If $a_{n+1} > a_{n+2-i}$, then we can scale up $\{a_0,a_1,\dots,a_{n}\}$ and scale down $\{a_{n+1},a_{n+2},\dots\}$ until $a_{n+1} = a_k$. It is easy to verify that this scaling operation will preserve the differential privacy constraint and decrease the cost.

Let $k^*$ be the smallest integer such that $a_{k^*} < a_{n+1}$. Note that by \eqref{eqn:less11} we have $ n+3-i \le k^* \le n$. Let $a_j$ be the biggest element in $\{a_{n+2},a_{n+3},\dots,a_{n+1+i}\}$. Due to the differential privacy constraint,  we have $\frac{a_j}{a_{n+1}} \le e^{\e}$. Then we can scale up $a_{k^*}$ and scale down $a_{n+1}$ until $a_{k^*} = a_{n+1}$  or $\frac{a_j}{a_{n+1}} = e^{\e}$. This operation will preserve the differential privacy constraint and decrease the cost. If after this scaling operation  $a_{k^*}$ is still bigger than $a_{n+1}$, then we can scale up $\{a_0,a_1,\dots,a_{n}\}$ and scale down $\{a_{n+1},a_{n+2},\dots\}$ until $a_{k^*} = a_{n+1}$. Due to the fact that $a_{n+1}$ is the smallest element in $\{a_{n+1},a_{n+2},\dots,a_{n+1+i}\}$, this scaling operation will preserve the differential privacy constraint and decrease the cost. Therefore, we will have $a_{n+1} \le a_{k^*}$.

Repeat the above steps for each $k \in {k^*+1, k^*+2, \dots, n}$ such that $a_k < a_{n+1}$. If we call the final probability distribution we obtained $\p_b$, we have $\p_b \in \sPsymi$, and the density sequence satisfying
\begin{align}
	b_0 \ge b_1 \ge b_2 \ge \cdots \ge b_n,
\end{align}
and $V(\p_b) \le V(\p_a)$.

Hence, among  $\sPsymi$ to minimize the cost we only need to consider probability distribution  with density sequence $\{a_0,a_1,a_2,\dots\}$ satisfying $a_0 \ge a_1 \ge a_2 \ge \cdots \ge a_n \ge a_{n+1}$.

Therefore, among all probability distributions in $\sPsymi$, to mininize the cost we only need to consider probability distributions with monotonically decreasing density sequence.

We conclude that
\begin{align}
	V^* = \inf_{\p \in \cup_{i=1}^{\infty} \pa}   V(\p) .
\end{align}

This completes the proof of Lemma \ref{lem:monotone}.
\end{IEEEproof}

%%%%%%%%%%%%%%%%%%%%%% third step, periodically decaying  %%%%%
\subsection{Step 3}

Next we show that among all symmetric piecewise constant probability density functions, we only need to consider those which are geometrically decaying.

More precisely, given positive integer $i$,
\begin{align}
	\pb \triangleq \{ \p | \p \in \pa, \; \text{and} \; \p \text{ has   density sequence}\;  \{a_0,a_1,\dots,a_n,\dots,\} \; \text{satisfying}  \frac{a_k}{a_{k+i}} = e^{\e}, \forall k \in \N \},
\end{align}
then
\begin{lemma}\label{lem:pd}
\begin{align}
	V^* = \inf_{ \p  \in \cup_{i=1}^{\infty} \pb}  V(\p).
\end{align}	
\end{lemma}

\begin{IEEEproof}
	Due to Lemma \ref{lem:monotone}, we only need to consider probability distributions with symmetric and piecewise constant probability density functions which are monotonically decreasing.
 	
 	We first show that given $\p_a \in \pa$ with density sequence $\{a_0,a_1,\dots,a_n,\dots,\}$, if $\frac{a_{0}}{a_{i}} < e^{\e}$, then we can construct a probability distributions $\p_b \in \pa$ with density sequence $\{b_0,b_1,\dots,b_n,\dots,\}$ such that $\frac{b_{0}}{b_{i}} = e^{\e}$ and
 	\begin{align}
 		V(\p_b) \le V(\p_a).
 	\end{align}

 	Define a new sequence $\{b_0,b_1,\dots,b_n,\dots\}$ by scaling up $a_0$ and scaling down $\{a_1,a_2,\dots\}$. More precisely, define $\{b_0,b_1,\dots,b_n,\dots\}$ via
 	\begin{align}
 	 	b_0 &= a_0 (1 + \delta), \\
 	 	b_k &= a_k ( 1 - \delta'), \forall \; k \ge 1,
 	 \end{align}
 for some $\delta >0$ and $0 < \delta' < 1$ such that
\begin{align}
	\frac{b_0}{b_i} &= e^{\e}, \\
	\sum_{k=0}^{+\infty} b_k \text{Vol}(A_i(k)) &= \sum_{k=0}^{+\infty} a_k \text{Vol}(A_i(k)) = 1.
\end{align}

So $\{b_0,b_1,\dots,b_n,\dots\}$ is a valid probability density sequence. Let $\p_b$ be the corresponding probability distribution. It is easy to check that $\p_b$ satisfies the differential privacy constraint, i.e.,
\begin{align}
 	\frac{b_k}{b_{k+i}} \le e^{\e}, \forall k \ge 0.
 \end{align}
Hence, $\p_b \in \pa$. Since $\cost(\|bx\|_1)$ is a monotonically increasing function of $\|\bx\|_1$, we have $V(\p_b) \le V(\p_a)$.

 	Therefore, for given $i \in \N$, we only need to consider $\p \in \pa$ with	 density sequence $\{a_0,a_1,\dots,a_n,\dots\}$ satisfying   $\frac{a_0}{a_i} = e^{\e}$.

 	Next, we argue that among all probability distributions  $\p \in \pa$ with	 density sequence $\{a_0,a_1,\dots,a_n,\dots,\}$ satisfying   $\frac{a_0}{a_i} = e^{\e}$,  we only need to consider those probability distributions with density sequence also  satisfying $\frac{a_1}{a_{i+1}} = e^{\e}$.

 	Given $\p_a \in \pa$ with	 density sequence $\{a_0,a_1,\dots,a_n,\dots\}$ satisfying   $\frac{a_0}{a_i} = e^{\e}$ and $\frac{a_1}{a_{i+1}} < e^{\e}$,  we can construct a new probability distribution $\p_b \in \pa$ with density sequence $\{b_0,b_1,\dots,b_n,\dots\}$  satisfying
 	\begin{align}
 	 	\frac{b_0}{b_i} &= e^{\e}, \\
 	 	\frac{b_1}{b_{i+1}} &= e^{\e},
 	 \end{align}
	and $V(\p_a) \ge V(\p_b)$.

	First, it is easy to see $a_1$ is strictly  less than $a_0$, since if $a_0 = a_1$, then $\frac{a_1}{a_{i+1}} = \frac{a_0}{a_{i+1}} \ge \frac{a_0}{a_i} = e^{\e}$. We can construct a new density sequence by increasing $a_1$ and decreasing $a_{i+1}$ to make $\frac{a_1}{a_{i+1}}$. More precisely, we define a new sequence $\{b_0,b_1,\dots,b_n,\dots\}$ as
	\begin{align}
		b_k &= a_k, \forall k \neq 1, k \neq i+1, \\
		b_1 &= a_1 (1 + \delta), \\
		b_{i+1} &= a_{i+1}(1 - \delta'),
	\end{align}
	where $\delta>0$ and $\delta'>0$ are chosen such that  $\frac{b_1}{b_{i+1}} = e^{\e}$ and
	\begin{align}
		\sum_{k=0}^{+\infty} b_k \text{Vol}(A_i(k)) &= \sum_{k=0}^{+\infty} a_k \text{Vol}(A_i(k)) = 1.
	\end{align}

	It is easy to verify that $\{b_0,b_1,\dots,b_n,\dots\}$ is a valid probability density sequence and the corresponding probability distribution $\p_b$ satisfies the differential privacy constraint \eqref{eqn:dpconstraintfinal}. Moreover, $V(\p_b) \le V(\p_a)$. Therefore, we only need to consider  $\p \in \pa$ with density sequences $\{a_0,a_1,\dots,a_n,\dots\}$  satisfying $\frac{a_0}{a_i} = e^{\e}$ and $\frac{a_1}{a_{i+1}} = e^{\e}$.

	% we can make this argument more rigorous. Basically,  we need to show that the new sequence must also be monotone decreasing. If $a^{1}_{i+1} \ge a^{1}_{i+1}$, then $\{a^{(1)}_k\}$ is a montone decreasing sequence, we are done. Suppose $a^{1}_{i+1} < a^{1}_{i+1}$, then we can rearrange the parts a_k for k>i+1 in decreasing order. And use the same delta technique. Since sum of a_k is finite. There are only finite number of k such that a_k is larger than a^(1)_{i+1}. So this change will terminate in finite number of steps. Therefore eventually we will get a monotone decreasing sequence.

	Use the same argument, we can show that  we only need to consider  $\p \in \pa$ with density sequences $\{a_0,a_1,\dots,a_n,\dots\}$  satisfying
	\begin{align}
		\frac{a_k}{a_{i+k}} &= e^{\e}, \forall k \ge 0.
	\end{align}

 	Therefore,
 	\begin{align}
		V^* = \inf_{ \p  \in \cup_{i=1}^{\infty} \pb}   V(\p).
	\end{align}

\end{IEEEproof}

Due to Lemma \ref{lem:pd}, we only need to consider  probability distribution with symmetric, monotonically decreasing, and geometrically decaying piecewise  constant probability density function. Because of the properties of symmetry and periodically (geometrically) decaying, for  this class of probability distributions, the probability density function over $\R^d$ is completely determined by the probability density function over the set $\{\bx \in \R^d | \|\bx\|_1 < \D\}$.

Next, we study what the optimal probability density function should be over the set $\{\bx \in \R^d | \|\bx\|_1 < \D\}$. It turns out that the optimal probability density function over the set $\{\bx \in \R^d | \|\bx\|_1 < \D\}$ is a step function. We use the following three steps to prove this result.

%%%%%%%%%%%%%%%%%%%%%%   4th step %%%%%%%%%%%%%%
\subsection{Step 4}
\begin{lemma}\label{lem:constline}
	Consider a probability distribution $\p_a \in \pb$ ($i \ge 2$) with density sequence $\{a_0,a_1,\dots, a_n,\dots\}$. Then there exists an integer $k(i)$ and a probability distribution $\p_b \in \pb$  with  density sequence $\{b_0,b_1,\dots, b_n,\dots\}$ such that
	\begin{align}
	 	b_0 &= b_1 = b_2 = \cdots = b_{k(i)}, \\
	 	\frac{b_0}{b_{i-1}} &= e^{\e},
	 \end{align}
	  and
	\begin{align}
		V(\p_b) \le  V(\p_a).
	\end{align}
\end{lemma}

\begin{IEEEproof}
	For $0 \le k \le i-1$, define
	\begin{align}
	 	w_k \triangleq \sum_{j=0}^{+\infty} e^{-j\e} \int \int \cdots \int_{(j+\frac{k}{i})\D \le \|\bx\|_1 < (j+\frac{k}{i})\D } \cost(\bx) d x_1 dx_2 \dots d x_d,
	 \end{align}
and
\begin{align}
	u_k \triangleq \sum_{j=0}^{+\infty} e^{-j\e} \text{Vol}(A_i(ji + k)).
\end{align}

Then the cost $V(\p_a) = \sum_{k=0}^{i-1} w_k a_k$, and the constraint on $a_k$ is that
\begin{align}
	a_0 &\ge a_1 \ge \cdots \ge a_{i-1}, \\
	a_0 &\le a_{i-1} e^{\e},\\
	\sum_{k=0}^{+\infty} u_k a_k &= 1.
\end{align}

Therefore, to mininize $V(\p)$ among all probability distributions $\p \in \pb$, we need to solve the following linear programming problem
\begin{align}
	\text{minimize}_{a_0,a_1,\dots,a_{i-1}} & \sum_{k=0}^{i-1} w_k a_k, \\
	\text{subject to}  \quad & a_0  \ge a_1 \ge \cdots \ge a_{i-1},\\
	& a_0  \le a_{i-1} e^{\e},\\
	& \sum_{k=0}^{+\infty} u_k a_k = 1.
\end{align}

Let
\begin{align}
	h_k \triangleq \frac{w_k}{u_k}. \label{eqn:h}
\end{align}

In the following we show that when $d=2$,  there exists an integer $k(i)$ such that
\begin{align}
  	h_0 &\ge h_1 \ge \cdots \ge h_{k(i)}, \label{eqn:ki1} \\
  	h_{k(i)} &\le h_{k(i)+1} \le \cdots \le h_{i-1}, \label{eqn:ki2}\\
  	h_0 &\le h_{i-1}. \label{eqn:ki3}
\end{align}

When $d=2$,
\begin{align}
	h_k &= \frac{w_k}{u_k} \\
		&= \frac{ \frac{4}{3} \frac{\D^3}{i^3}  \sum_{j=0}^{+\infty} e^{-j\e} (1+3(ji+k)+3(ij+k)^2}{ 2 \frac{\D^2}{i^2} \sum_{j=0}^{+\infty} e^{-j\e} (1+2(ji+k)) } \\
		%&= \frac{2}{3} \frac{\D}{i} \frac{3i^2 \frac{b^2+b}{(1-b)^3} + (6ik+3i)\frac{b}{(1-b)^2} + (1+3j+3j^2)\frac{1}{1-b} }{ \frac{1+2j}{1-b} + \frac{2ib}{(1-b)^2}}.
		&= \frac{2}{3} \frac{\D}{i} \frac{ 3i^2 c_2 + (6ik+3i)c_1 + (1+3k+3k^2) c_0 }{(1+2k)c_0 + 2i c_1}.
\end{align}

Let $g(k) = \triangleq \frac{ 3i^2 c_2 + (6ik+3i)c_1 + (1+3k+3k^2) c_0 }{(1+2k)c_0 + 2i c_1}$. It is easy to compute the derivative of $g(k)$ with respect to $k$:
\begin{align}
	g'(k) = \frac{  6c_0^2 k^2 + 6c_0^2 k + c_0^2 + 12 c_0 c_1 i k + 6c_0 c_1 i - 6 c_2 c_0 i^2 + 12 c_1^2 i^2}{ ( (1+2k)c_0 + 2i c_1 )^2 }.
\end{align}

Note that the numerator of $g'(k)$ is an increasing function of $k$, and
\begin{align}
	g'(0) &= c_0^2 + 6c_0 c_1 i - 6 c_2 c_0 i^2 + 12 c_1^2 i^2 \\
	&= \frac{ b(6i^2-6i+1) -1 }{(b-1)^3} <0,
\end{align}
for sufficiently large $i$, and
\begin{align}
	g'(i-1) &= \frac{6i^2 - 6i + 1 - b }{(1-b)^3} >0.
\end{align}

Therefore, $h_k$ first increases as $k$ increases, and then decreases as $k$ increases to $i-1$. Hence, there exists an integer $k(i)$ such that \eqref{eqn:ki1} and \eqref{eqn:ki2} hold.

Next we compare $h_{i-1}$ and $h_0$:
\begin{align}
	h_{i-1} - h_0 &= \frac{w_{i-1}}{u_{i-1}} - \frac{w_0}{u_0} \\
	&= \frac{2}{3} \frac{\D}{i} \frac{ (3i-2)(b-1)^2 (i-1) }{ (2bi-b+1) (b+2i-1)  } >0.
\end{align}
Hence, \eqref{eqn:ki3} also holds.

Now we are ready to prove Lemma \ref{lem:constline}.

Suppose $a_{k(i)} < a_{k(i)-1}$.  We can scale up $a_{k(i)}$  and scale down $a_{k(i)-1}$ to make $a_{k(i)} = a_{k(i)-1}$. Since $h_{k(i)} \le h_{k(i)-1}$, i.e., $\frac{w_{k(i)}}{u_{k(i)}} \le \frac{w_{k(i)-1}}{u_{k(i)-1}}$, this scaling operation will not increase the cost $V(\p_a)$. Now we have $a_{k(i)} = a_{k(i)-1}$.

Suppose  $a_{k(i)} = a_{k(i)-1} < a_{k(i)-2}$. Then we can scale up $a_{k(i)} $ and $a_{k(i)-1}$, and scale down $a_{k(i)-2}$ to make $a_{k(i)} = a_{k(i)-1} = a_{k(i)-2}$. Since $h_{k(i)} \le h_{k(i)-1} \le h_{k(i)-2}$, this scaling operation will not increase the cost $V(\p_a)$. Now we have $a_{k(i)} = a_{k(i)-1} = a_{k(i)-2}$.

After $k(i)$ steps of these scaling operations, we can make $a_0 = a_1 = \cdots = a_{k(i)}$, and this will not increase the cost $V(\p_a)$.

Finally, if $\frac{a_0}{a_{i-1}} < e^{\e}$, we can scale up $a_0, a_1, \dots, a_{k(i)}$, and scale down $a_{i-1}$ to make $\frac{a_0}{a_{i-1}} = e^{\e}$. Since $h_{i-1} \ge h_0 \ge h_1 \ge \cdots \ge h_{k(i)}$, this scaling operation will not increase the cost $V(\p_a)$.

Let $\p_b$ be the probability distribution we obtained after the $k(i)+1$ steps of scaling operations. Then $\p_b \in \pb$, and its density sequence $\{b_0,b_1,\dots, b_n,\dots\}$ satisfies
	\begin{align}
	 	b_0 &= b_1 = b_2 = \cdots = b_{k(i)}, \\
	 	\frac{b_0}{b_{i-1}} &= e^{\e},
	 \end{align}
	  and
	\begin{align}
		V(\p_b) \le  V(\p_a).
	\end{align}

This completes the proof of  Lemma \ref{lem:constline}.

\end{IEEEproof}

Therefore, due to Lemma \ref{lem:constline}, for sufficiently large  $i$, we only need to consider probability distributions $\p \in \pb$ with density sequence  $\{a_0,a_1,\dots, a_n,\dots\}$ satisfying
\begin{align}
	a_0 &= a_1 = a_2 = \cdots = a_{k(i)},  \label{eqn:twosteps1}\\
	\frac{b_0}{b_{i-1}} &= e^{\e}. \label{eqn:twosteps2}
\end{align}

More precisely, define
\begin{align}
	\pc = \{ \p \in \pb | \p \; \text{has density sequence} \; \{a_0,a_1,\dots, a_n,\dots\} \; \text{satisfying}  \;  \eqref{eqn:twosteps1}  \; \text{and} \;  \eqref{eqn:twosteps2}  \}.
\end{align}
Then due to Lemma \ref{lem:constline},
\begin{lemma}\label{lem:ratiolem}
	\begin{align}
		V^* = \inf_{\p \in \cup_{i=3}^{\infty}\pc}  V(\p).
	\end{align}
\end{lemma}

%%%%%%%%%%%%%%%%%%%%%%%%%%%%  Step 7 %%%%%%%%%%%%%%%%
% \subsection{Step 7}

Next, we argue that  for each probability distribution  $\p \in \pc$  ($i \ge 3$) with density sequence $\{a_0,a_1,\dots, a_n,\dots\}$, we can assume that there exists an integer $k(i) + 1 \le k \le (i-2)$, such that
\begin{align}
	a_j &= a_0, \forall 0 \le j <k, \label{eqn:property1} \\
	a_j &= a_{i-1}, \forall k< j <i. \label{eqn:property2}
\end{align}

More precisely,
\begin{lemma}\label{lem:binary}
		Consider a probability distribution $\p_a \in \pc$ ($i \ge 3$) with density sequence $\{a_0,a_1,\dots,a_n,\dots\}$. Then there exists a probability distribution $\p_b \in \pc$ with density sequence $\{b_0,b_1,\dots,b_n,\dots\}$ such that there exists an integer $k(i) + 1 \le k \le (i-2)$ with
		\begin{align}
		b_j &= a_0, \forall \;  0 \le j <k, \label{eqn:binary1}\\
		b_j &= a_{i-1}, \forall \; k< j <i, \label{eqn:binary2}
		\end{align}
and
	\begin{align}
		V (\p_b) \le  V (\p_a). \label{eqn:binary3}
	\end{align}
\end{lemma}

\begin{IEEEproof}
	If there exists an integer $k(i) + 1 \le k \le (i-2)$ such that
		\begin{align}
		a_j &= a_0, \forall \; 0 \le j <k, \\
		a_j &= a_{i-1}, \forall \; k< j <i,
		\end{align}
then we can set $\p_b = \p_a$.

Otherwise, let $k_1$ be the smallest integer in $\{k(i)+1,k(i)+2,\dots,i-1\}$ such that
\begin{align}
	a_{k_1} \neq a_0,
\end{align}
and let $k_2$ be the biggest integer in $\{k(i)+1,k(i)+2,\dots,i-1\}$ such that
\begin{align}
	a_{k_2} \neq a_{i-1}.
\end{align}

It is easy to see that $k_1 \neq k_2$. Then we can scale up $a_{k_1}$ and scale down $a_{k_2}$ simultaneously until either $a_{k_1} = a_0$ or $a_{k_2} = a_{i-1}$. Since $h_k \triangleq \frac{w_k}{u_k}$ is an increasing function of $k$ when $k > k(i)$, and $k(i)< k_1 < k_2$, this scaling operation will not increase the cost.

After this scaling operation we can update $k_1$ and $k_2$, and either $k_1$ is increased by one or $k_2$ is decreased by one.

Therefore, continue in this way, and finally we will obtain a probability distribution $\p_b \in \pc$ with density sequence $\{b_0,b_1,\dots, b_n,\dots\}$ such that \eqref{eqn:binary1}, \eqref{eqn:binary2} and \eqref{eqn:binary3} hold.

This completes the proof.

\end{IEEEproof}

Define
\begin{align}
	\pd = \{ \p \in \pc \ |\ \p \; \text{has density sequence} \; \{a_0,a_1,\dots,a_n,\dots\} \;  \text{satisfying} \eqref{eqn:binary1} \; \text{and} \; \eqref{eqn:binary2} \; \text{for some}\; k(i) < k \le (i-2)  \}.
\end{align}

Then due to Lemma \ref{lem:binary},
\begin{lemma}\label{lem:stepfunc}
	\begin{align}
		V^* = \inf_{\p \in \cup_{i=3}^{\infty}\pd} V(\p).
	\end{align}
\end{lemma}

As $i \to \infty$, the probability density function of $\p \in \pc$ will converge to a multidimensional staircase function. Therefore, for $d=2$ and the cost function $\loss(\bx) = \|\bx\|_1, \forall \bx \in \R^2$, then
\begin{align}
	\inf_{\p \in \sP}  \int  \int_{\R^2} \loss(\bx) \p (d x_1 d x_2 )  = \inf_{\gamma \in [0,1]}  \int  \int_{ \R^2} \loss(\bx)     f_{\gamma}(\bx)  d x_1 d x_2  .
\end{align}

This completes the proof of Theorem \ref{thm:main}.

\bibliographystyle{IEEEtran}

\bibliography{reference}

\end{document}